# Air Stable Doping and Intrinsic Mobility Enhancement in Monolayer Molybdenum Disulfide by Amorphous Titanium Suboxide Encapsulation

Amritesh Rai[1] [*], Amithraj Valsaraj[1], Hema C.P. Movva[1], Anupam Roy[1], Rudresh Ghosh[1], Sushant Sonde[1], Sangwoo Kang[1], Jiwon Chang[2], Tanuj Trivedi[1], Rik Dey[1], Samaresh Guchhait[1], Stefano Larentis[1], Leonard F. Register[1], Emanuel Tutuc[1], and Sanjay K. Banerjee[1]

[1] Microelectronics Research Center & Department of Electrical and Computer Engineering, The University of Texas at Austin, Austin, Texas 78758, USA

[2] SEMATECH, 257 Fuller Rd #2200, Albany, NY-12203, USA

[*] Correspondence e-mail: amritesh557@utexas.edu

**Abstract:** To reduce Schottky-barrier-induced contact and access resistance, and the impact of charged impurity and phonon scattering on mobility in devices based on 2D transition metal dichalcogenides (TMDs), considerable effort has been put into exploring various doping techniques and dielectric engineering using high-κ oxides, respectively. The goal of this work is to demonstrate a high-κ dielectric that serves as an effective n-type charge transfer dopant on monolayer (ML) molybdenum disulfide ($MoS_2$). Utilizing amorphous titanium suboxide (ATO) as the 'high-κ dopant', we achieved a contact resistance of ~ 180 Ω·µm which is the lowest reported value for ML $MoS_2$. An ON current as high as 240 µA/µm and field effect mobility as high as 83 $cm^2$/V-s were realized using this doping technique. Moreover, intrinsic mobility as high as 102 $cm^2$/V-s at 300 K and 501 $cm^2$/V-s at 77 K were achieved after ATO encapsulation which are among the highest mobility values reported on ML $MoS_2$. We also analyzed the doping effect of ATO films on ML $MoS_2$, a phenomenon which is absent when stoichiometric $TiO_2$ is used, using *ab initio* density functional theory (DFT) calculations which shows excellent agreement with our experimental findings. Based on the interfacial-oxygen-vacancy mediated doping as seen in the case of high-κ ATO – ML $MoS_2$, we propose a mechanism for the mobility enhancement effect observed in TMD-based devices after encapsulation in a high-κ dielectric environment.

Research on two-dimensional (2D) layered materials has been on the rise ever since the isolation of graphene in 2004.[1] Despite its remarkable electrical, optical and mechanical properties, the lack of a sizeable band-gap hinders the use of graphene in next generation digital electronic applications.[2] Alternatively, semiconducting layered transition metal dichalcogenides (TMDs) can circumvent this issue owing to their band-gaps, thereby paving the way for the realization of digital logic circuits utilizing 2D layered materials. Semiconducting TMDs offer unique advantages when incorporated into field effect transistors (FETs) in their atomically thin form. Their high transparency, mechanical flexibility and mobility values higher than organic, amorphous or polycrystalline semiconductors make them ideal candidates for use in future transparent and flexible electronics.[3] Moreover, their ultrathin body allows for excellent electrostatic gate control of the channel even at aggressively scaled channel lengths, thereby making them less susceptible to detrimental short channel effects.[2,3,4,5] Molybdenum disulfide ($MoS_2$) has emerged as the most popular and extensively studied semiconducting TMD over the past few years due to its natural abundance, and excellent environmental stability. The potential of $MoS_2$ has been demonstrated in various electronic and optoelectronic device applications including FETs,[6] photovoltaics,[7] photodetectors,[8] sensors,[9] as well as in more complicated logic circuits such as inverters,[10] memory cells,[11] ring oscillators,[12] etc. However, FETs made from $MoS_2$ and other TMDs suffer from a large contact resistance ($R_C$) due to the Schottky barriers formed at the metal-TMD interface.[13,14] Moreover, extrinsic mobility limiting factors such as charged impurity scattering adversely affect the device performance.[15] Realization of ohmic contacts and minimization of mobility limiting factors is necessary to harvest the potential of $MoS_2$ in future nanoelectronic applications, especially at aggressively scaled channel lengths.

In order to alleviate the large $R_C$ in $MoS_2$ FETs, insightful work has been done by several groups over the past few years. Das et al. demonstrated the use of low work function scandium as an efficient electron injector into the conduction band of $MoS_2$.[16] N-type charge transfer dopants such as potassium ions (K)[17] and polyethyleneimine (PEI)[18] have been utilized, although these doping reagents are unstable in ambient conditions. The use of graphene-metal hetero-contacts[19] and air stable doping via benzyl viologen (BV)[20] were shown to be effective strategies, but the $R_C$ values were still greater than 1 kΩ·µm and only moderate channel mobilities were achieved. More recently, the use of phase engineered contacts[21] and chloride doping[22,23] were demonstrated on $MoS_2$ with promising results. However, the stability of the phase engineered contacts under high-performance device operation is still unknown.[21] Furthermore, the chloride doping mechanism is unclear and it is speculated that the doping occurs due to filling of the naturally occurring sulfur vacancies in $MoS_2$ by chlorine atoms.[23] Besides doping and the corresponding $R_C$ reduction, considerable effort has also been put into dielectric engineering utilizing high dielectric constant (high-κ) materials to reduce the scattering of carriers in $MoS_2$ devices. Although several high-κ dielectrics have been investigated, atomic layer deposition (ALD) of alumina and hafnia have been the most common choices.[6,10,15,24,25,26,27]

In this letter, we demonstrate an air stable, self-encapsulating, n-type charge transfer doping technique on ML $MoS_2$ utilizing amorphous titanium sub-oxide (ATO) thin films. The ATO can be solution processed in the form of a sol-gel precursor and its application involves a simple spin-coating process, thereby making this approach extremely facile and easily scalable in contrast to the phase engineering or chloride doping schemes which require several hours of treatment with their respective chemical reagents.[21,22,23] Utilizing this technique, we achieved a very low $R_C$ of ~ 180 Ω·μm on ML $MoS_2$, which compares favorably to the $R_C$ values obtained on 2-3 layer $MoS_2$ with phase engineered contacts,[21] and is ~ 2.5 times lower than the $R_C$ reported on chloride-doped multilayer $MoS_2$ FETs.[23] An ON-current as high as 240 μA/μm was achieved for a 450 nm channel length (L) back gated FET, with an oxide thickness ($t_{ox}$) of 93 nm, at a drain-to-source voltage ($V_{DS}$) of 2 V and back-gate overdrive voltage ($V_{BG}$ - $V_T$) of 70 V. Field-effect mobilities ($\mu_{FE}$) as high as 83 $cm^2$/V-s and intrinsic mobilities ($\mu_{int}$) as high as 102 $cm^2$/V-s were achieved on ML $MoS_2$ devices at room temperature (RT) upon ATO encapsulation. Temperature-dependent measurements revealed enhanced intrinsic mobilities approaching 501 $cm^2$/V-s in ATO encapsulated ML $MoS_2$ at 77 K. Density functional theory (DFT) analysis was performed to gain further insight into the doping mechanism of ATO films on ML $MoS_2$.

The mechanism of charge transfer doping is particularly attractive for ultrathin layered materials since it does not involve any substantial distortion of the 2D crystal lattice.[20] Several charge transfer doping techniques that were previously demonstrated on carbon based nanomaterials were also successfully demonstrated on $MoS_2$.[17,18,20] Similarly, we investigate the effects of high-κ ATO thin films on $MoS_2$ which serves as an n-type charge transfer dopant. For the purpose of this experiment, only ML $MoS_2$ flakes were considered. Details on materials and device fabrication methods can be found in supporting information S1, while the characterization tools and techniques are described in supporting information S2. ATO thin films were deposited on MoS$_2$ FETs by spin-coating at 3000 rpm and subsequent baking of an ATO sol-gel precursor solution [see supporting information S3 for its preparation method] at 90$^0$ C on a hot plate for 15 minutes to dry the residual solvent and convert the precursor solution into ATO through hydrolysis. ATO thin films obtained using this process were reported to have band-gaps of ~ 3.7 eV[28] and ~ 3.9 eV[29] corresponding to a Ti:O ratio of 1:1.34,[28] and 1:1.59,[29] respectively. The amorphous nature of these films and their large band-gaps have been confirmed in literature by X-ray diffraction (XRD) and optical absorption measurements, respectively.[28,29] The Ti:O ratio in our films was estimated to be ~ 1:1.5 from the XPS data [supporting information S4] confirming the oxygen deficiency. Also, from the reported band-gaps for ATO films with different Ti:O ratios,[28,29] the band-gap of our films can be estimated to be between 3.7 eV and 3.9 eV. Therefore, ATO can effectively be regarded as a wide band-gap amorphous oxide semiconductor. As TiO$_2$ can serve as channel for n-type thin film transistors,[30,31,32] it is important to first rule any parallel conduction paths that can be added to the MoS$_2$ channel by the encapsulating ATO layer. To test for possible conduction through the ATO film, a set of control devices without the MoS$_2$ channel were fabricated in exactly the same manner as the actual devices. No conduction was observed through the as-formed ATO layer even under higher biasing conditions (both back-gate and drain) than what was used in actual devices. Hence, the ATO films in our case were found to be completely insulating.

Figure 1(a) shows an image of the as-prepared ATO sol-gel precursor solution with a concentration of ~ 85 mg/ml. Figure 1(b) shows a schematic of the chemistry responsible for the formation of ATO from titanium isopropoxide, its precursor molecules. A schematic of a back-gated FET encapsulated by ATO is illustrated in Figure 1 (c). The doping of MoS$_2$ leads to changes in its Raman and photoluminescence (PL) spectra. Figure 1(d) compares the normalized Raman spectra of an as-exfoliated ML MoS$_2$ flake (blue) to that of the same flake after encapsulation by ATO (red). The peak positions of the out-of-plane A$_{1g}$ and in-plane E$_{2g}^1$ peaks for the bare ML MoS$_2$ are at 402.0 cm$^{-1}$ and 383.0 cm$^{-1}$, respectively, corresponding to a peak separation of 19 cm$^{-1}$. This peak separation is characteristic of ML MoS$_2$.[33] Upon encapsulation with ATO, the E$_{2g}^1$ peak positon and peak full-width half-maximum (FWHM) remain relatively unchanged. On the other hand, the A$_{1g}$ peak shows a distinct broadening with its FWHM increasing from 6.6 cm$^{-1}$ to 8.1 cm$^{-1}$, as well as a red shift from 402.0 cm$^{-1}$ to 399.6 cm$^{-1}$. This red shift and peak broadening of the A$_{1g}$ Raman mode are characteristic of doped MoS$_2$ and have been observed in previous doping studies.[20] Figure 1(e) compares the PL spectra of a ML MoS$_2$ flake before (blue) and after (red) encapsulation with ATO. Before encapsulation, the peak position of the A exciton is at 1.86 eV, consistent with reported values for ML MoS$_2$.[34] Upon ATO encapsulation, the A exciton peak shows a decrease in intensity and a red shift of 16 meV, which can be attributed to the formation of negatively charged trions from excitons as a result of the increased electron concentration.[35,36] The pronounced changes in the Raman and PL spectra of ML MoS$_2$ upon ATO encapsulation clearly indicate the n-type doping effects of ATO on MoS$_2$.

The transfer characteristics of a representative back-gated MoS$_2$ FET at V$_{DS}$ = 1 V, before and after ATO encapsulation, as well as after one month of exposure to ambient conditions are shown in Figure 2(a). All FETs were fabricated on 93 nm SiO$_2$/n$^{++}$ Si substrates. The transfer curve before doping (blue) indicates a strong electrostatic gate control over the channel with a threshold voltage (V$_T$) of 7 V, extracted from the linear region of the transfer characteristics, and an I$_{ON}$/I$_{OFF}$ ratio up to 10$^8$. Upon encapsulation with ATO, the gate modulation is significantly reduced (red curve), and the V$_T$ shifts to -25 V. This large negative V$_T$ shift is indicative of the n-doping effect of ATO. The 2D sheet electron concentration (n$_{2D}$) after ATO doping can be estimated as n$_{2D}$ = (C$_{OX}$ |ΔV$_T$|)/q, where q is the electron charge, C$_{OX}$ = 3.71 × 10$^{-8}$ F/cm$^2$ is the gate oxide capacitance, and ΔV$_T$ = -32 V is the shift in threshold voltage right after doping. The extracted value of n$_{2D}$ for this device upon doping was 7.4 × 10$^{12}$ cm$^{-2}$. Previous doping studies on MoS$_2$ utilizing K ions[17] and benzyl viologen[20] reported n$_{2D}$ values of 1 × 10$^{13}$ cm$^{-2}$ and 1.2 × 10$^{13}$ cm$^{-2}$, respectively. The n$_{2D}$ value as a result of ATO doping is slightly lower in our case, however, it should be noted that our experiments used ML flakes unlike previous studies that utilized multilayer flakes. The long term air stability of encapsulated ATO doping is evident from the electrical data as even after thirty days of exposure to ambient conditions, the device shows similar ON-currents, a weak gate modulation and has an n$_{2D}$ = 3.7 × 10$^{12}$ cm$^{-2}$ (green curve). Moreover, by virtue of being self-encapsulating, ATO films protect the underlying MoS$_2$ channel from the degrading effects of atmospheric adsorbates. However, there is slight performance degradation after extended ambient exposure in ATO encapsulated devices [supporting information S5].

The inset of Figure 2(a) shows the transfer characteristics of the same device at larger gate and drain biases following the ATO encapsulation. The 450 nm channel length device could be switched off to a moderate extent ($I_{ON}/I_{OFF}$ = 4 x 10$^3$, subthreshold swing = 1.6 V/decade) at large negative gate biases even though $V_{DS}$ was as high as 2 V. Further optimization and control over the starting concentrations of the ATO precursor solution or realization of top gated devices with ATO encapsulated S-D access regions would help yield an ideal balance between $I_{ON}/I_{OFF}$ ratio and high saturation ON-currents. Figure 2(b) shows the output characteristics of the same device as in Figure 2(a). After ATO encapsulation, the ON current of the ML device at $V_{DS}$ = 1 V and $V_{BG}$ = 25 V is 144 µA/µm which is 2.5 times greater than the corresponding value for the undoped device. The inset of Figure 2(b) shows the output characteristics of the same device subject to larger biasing conditions. At a $V_{DS}$ of 2 V and $V_{BG}$ of 45 V, the ON current is as high as 240 µA/µm showing the onset of current saturation at large positive gate and drain biases. Our ATO-doped ML MoS$_2$ FET with an ON current of 240 µA/µm compares well with the highest drain current to date on chloride-doped multilayer MoS$_2$ FETs,[22,23] taking into account the fact that the channel length in our case was 4.5 times larger and the device was made on a ML flake.

In order to quantify the effect of ATO doping on the electrical contact between the metal (Ag) and the ML MoS$_2$, a transfer length method (TLM) analysis was carried out. A suitable large area ML flake was identified, upon which a set of contacts were fabricated with different channel lengths as shown in the inset of Figure 3(b). The basic equation underlying the TLM analysis can be written as $R_{TOTAL} = (R_{SH}L)/W + 2R_C$, where $R_{TOTAL}$ is the total measured resistance of a channel between two contacts, $R_{SH}$ is the sheet resistance of the channel, L and W are the channel's length and width, respectively, and $R_C$ is the contact resistance. By fitting a plot of ($R_{TOTAL} \cdot W$) as a function of L, key parameters such as $R_{SH}$, $R_C$ and transfer length ($L_T$) can be extracted. Figure 3(a) shows the total resistance, measured at a $V_{BG}$ of 25 V and $V_{DS}$ of 0.1 V, as a function of L before (blue) and after (red) ATO encapsulation. From a linear fit to the measured resistances before doping, an $R_{SH}$ of 20.1 kΩ/□, $R_C$ of 2.9 kΩ·µm and a transfer length ($L_T$) of 145 nm were extracted. Fitting the measured resistances after ATO encapsulation, we extracted an $R_{SH}$ of 12.4 kΩ/□, $R_C$ of ~ 180 Ω·µm (inset of Figure 3(a)) and an $L_T$ of 15 nm. This significant reduction in $R_{SH}$, $R_C$ and $L_T$ upon ATO encapsulation reflects the efficacy of this doping technique. This is the lowest reported $R_C$ value among all previous n-type doping studies on

MoS$_2$[17,18,20] and compares well with the recently reported record low R$_C$ value (~ 80 Ω·μm at a V$_{BG}$ of 30 V) on MoS$_2$ with phase engineered contacts.[21]

Figure 3(b) shows the extracted R$_C$ values plotted as a function of V$_{BG}$ before and after ATO encapsulation. For the undoped case (blue curve), the R$_C$ shows a strong dependence on gate bias and increases exponentially at negative gate biases due to the large Schottky barriers present at the contacts. On the other hand, for the ATO doped case (red curve), the R$_C$ is fairly independent of the applied gate bias for V$_{BG}$ > -10 V. This results from the substantial thinning of the Schottky barrier width as a consequence of heavy doping at the contact regions. This Schottky barrier thinning effect is also apparent in the transfer characteristics temperature dependence, and in the output characteristics measured at 77 K of a back-gated ML MoS$_2$ FET after ATO doping [supporting information S6]. Thus, in the ATO-doped ML MoS$_2$ devices, the effective Schottky barriers are significantly reduced even though the doping occurs along the contact edges as opposed to directly underneath the contacts. We note that this ATO doping effect on ML MoS$_2$ is absent when stoichiometric TiO$_2$ is used, as demonstrated previously in the case of graphene.[29,37] This was verified by depositing TiO$_2$ on back-gated ML MoS$_2$ FETs utilizing a recently demonstrated technique[38] [supporting information S7].

To gain further insight into the doping mechanism of MoS$_2$ by ATO, an *ab initio* DFT analysis was carried out to study the effects of both a Ti-rich and an O-rich interface of an underlying TiO$_2$ slab on the electronic structure of ML MoS$_2$ via band-structure and atom-projected density-of-states (AP-DOS) calculations. The DFT simulation was performed using the Vienna Ab initio Simulation Package (VASP)[39,40] and exact details of the methodology employed here are described elsewhere.[41] Briefly, our simulations were performed by constructing a supercell of ML MoS$_2$ on an approximately 1 nm thick TiO$_2$ slab. Atomic relaxation was performed within a rectangular supercell (*a* = 9.366 Å, *b* = 5.407 Å) chosen to reduce the lattice mismatch between ML MoS$_2$ and rutile-TiO$_2$ as shown in Figure 4(a). The rutile phase was chosen for the simulation since it is the most common natural form of TiO$_2$.[42] As stated before, we consider two possible terminations for the TiO$_2$ slab, a Ti-rich TiO$_2$ slab and an O-rich TiO$_2$ slab. For the Ti-rich TiO$_2$ case, the surface O-atoms were removed from the supercell corresponding to an O-vacancy density of 7.896×10$^{14}$/cm$^2$ in order to mimic the ATO structure with interfacial O-vacancies. In these 0 K simulations, the highest occupied state corresponds to the 0 eV reference energy. Figure 4(b) shows the band structure of ML MoS$_2$ on a Ti-rich TiO$_2$ slab depicting occupied conduction bands below the Fermi level leading to a system that appears metallic. From the corresponding AP-DOS plot shown at the right, we can observe that the occupied conduction bands can be attributed to Ti, Mo and S atom states implying that the additional states introduced by the Ti atoms appear near the conduction band states of ML MoS$_2$. For the composite MoS$_2$-TiO$_2$ system, this phenomenon can be interpreted as a transfer of electrons into the lower conduction-band-edge of the ML MoS$_2$ layer analogous to modulation doping. In contrast, this phenomenon is absent in the case of ML MoS$_2$ on the O-rich TiO$_2$ slab as depicted in Figure 4(c) wherein, we have an ideal TiO$_2$ surface without any O-vacancies in the supercell. Here, the Fermi level is pinned at the valence band edge and the conduction band states remain unoccupied. Hence, our theoretical findings are in excellent agreement with our experimental results. It is to be noted that in the band-structures depicted in Figures 4(b) and (c), the conduction band minima and the valence band maxima are located at the Γ point as opposed to the K point for ML MoS$_2$. This is because using a bigger supercell in the DFT simulations results in the corresponding brillouin zone being smaller and, hence, the K point folds into the Γ point.[41]

An added advantage of using high-κ ATO as a self-encapsulating dopant is the intrinsic mobility enhancement of ML MoS$_2$ as extracted from four-point back-gated devices which exclude contact resistance effects. High-κ dielectric engineering, using ALD deposited hafnia and alumina, has been used widely on MoS$_2$ and other TMDs.[6,10,16,24,25,43] Although the exact mechanism is still unclear, it is believed that the presence of a high-κ environment enhances the carrier mobility by 'screening' the Coulomb interactions with charged impurities, as well as by quenching the homopolar phonon modes of MoS$_2$.[15,44,45] Although ATO films have been shown to have a κ value ranging between 70 – 120,[46] the κ value of our solution-processed ATO films was extracted to be ~ 10 from capacitance-voltage measurements, a value comparable to the κ-values reported for alumina and hafnia.[47] Figure 5(a) shows the measured four-point conductance (G$_{4\text{-pt}}$) as a function of V$_{BG}$ – V$_T$ for a ML MoS$_2$ device (shown in the inset with the flake outlined at its edges) before and after ATO encapsulation. The four left-most contacts of the device (a, b, c and d) were used for the four point measurement which was done at RT. Current was passed between the outer two contacts (a, d) while the inner two contacts (b, c) served as the voltage probes. A marked difference exists between the slopes of the curves from the bare device (blue) and after its encapsulation in ATO (red). Intrinsic mobility ($\mu_{int}$) was calculated using the expression $\mu_{int} =$ (L/W) (1/$C_{OX}$) (d$G_{4-pt}$/d$V_{BG}|_{max}$) where L = 1.45 µm and W = 1.42 µm are the length and width of the active region bounded by contacts b and c, respectively, $C_{OX}$ is the geometric oxide capacitance, and (d$G_{4-pt}$/d$V_{BG}|_{max}$) is the maximum slope of the four point conductance curves as marked by dashed light-green lines in the figure. For the bare ML $MoS_2$, we extracted a $\mu_{int}$ of 48 $cm^2$/V-s at RT whereas after encapsulation, $\mu_{int}$ increased to 102 $cm^2$/V-s (~ 2X improvement). This value is among the highest intrinsic mobilities reported for ML $MoS_2$ at RT and comes close to the calculated RT phonon-limited mobility of 130 $cm^2$/V-s, a more realistic estimation in which the effect of inter-valley scattering between the K and Q valleys, separated from each other in energy by just 70 meV, was also considered.[48] Though this was the best RT intrinsic mobility enhancement we observed upon ATO encapsulation (> 2X improvement), the effect itself was observed in five other four point devices. The two-point $\mu_{FE}$ measured between contacts 'd' and 'e' (L = 0.46 µm, W = 1.42 µm) at a $V_{DS}$ of 100mV before and after encapsulation was 24 $cm^2$/V-s and 83 $cm^2$/V-s, respectively, showing a > 3X improvement [supporting information S8]. Comparing the two point $\mu_{FE}$ of this device with the four point $\mu_{int}$ of the parent $MoS_2$ flake, we see that the ratio $\mu_{int}/\mu_{FE}$ decreases from 2.02 before ATO encapsulation to 1.23 after ATO encapsulation implying that the two-point $\mu_{FE}$ of this device approaches the four-point $\mu_{int}$ of the parent flake due to the doping by ATO.

Figure 5(b) shows the maximum four-point intrinsic mobility of another ML $MoS_2$ device (shown in the inset) as a function of temperature. The length and width of the active region are 2.3 µm, and 2.5 µm (flake width), respectively. Before ATO encapsulation (blue), the intrinsic mobility varies from 30 $cm^2$/V-s at RT to 285 $cm^2$/V-s at 77 K. After ATO encapsulation (red), the values range from 52 $cm^2$/V-s at RT to 501 $cm^2$/V-s at 77 K following a similar trend. This value of 501 $cm^2$/V-s in ATO-encapsulated $MoS_2$ is among the highest intrinsic mobilities reported till date on ML $MoS_2$ at 77 K, and compares well with the recent work on ultra-high mobility $MoS_2$ that is encapsulated in hexagonal boron nitride and contacted by graphene.[49] Although this mobility enhancement may be attributed to the high-κ nature of the encapsulating ATO, we know that the $n_{2D}$ in the $MoS_2$ channel is increased as the high-κ ATO film dopes the $MoS_2$ owing to its interfacial oxygen vacancies. Increased carrier densities in a non-degenerate 2D channel aids in enhancing the carrier mobility by screening the charged impurities, as has been demonstrated both theoretically[44] and experimentally[50] in ML $MoS_2$. Furthermore, the increased electron concentration also serves to soften the homopolar phonons of $MoS_2$ as evident from the red shift and broadening of the out-of-plane $A_{1g}$ Raman mode of ML $MoS_2$ upon ATO encapsulation. Our results, therefore, give important insight into the mechanism of mobility enhancement in $MoS_2$ devices effected by high-κ dielectrics. In light of our ATO – $MoS_2$ results, it is plausible that this doping effect can be caused by other high-κ dielectrics, such as ALD deposited alumina or hafnia, if they have inherent oxygen vacancies at their interfaces with $MoS_2$. Given the amorphous nature of the ALD grown high-κ dielectrics, it is highly possible that oxygen vacancies exist in their structure. In fact, our recent investigation[41] reveals that interfacial oxygen vacancies in alumina or hafnia lead to the creation of donor states near the conduction band of $MoS_2$. These donor states originate from the uncompensated aluminum and hafnium atoms at the high-κ – $MoS_2$ interface, much akin to our case of uncompensated titanium atoms at the ATO – $MoS_2$ interface, resulting in n-type doping of the ML $MoS_2$ channel. On the other hand when the alumina or hafnia is perfectly stoichiometric, no doping effect is observed. Therefore, we propose that this interfacial-oxygen-vacancy mediated doping effect plays a prominent role in enhancing both the intrinsic and field-effect mobility in high-κ encapsulated TMD devices. Upon high-κ encapsulation, there would be an increase in the $n_{2D}$ of the TMD channel even before the application of external gate or drain biases, and this increased $n_{2D}$ would screen out the charged impurities, suppress the homopolar phonons and reduce the effective Schottky barrier at the contacts to a greater extent than in bare devices. Hence, when the external biases are applied, the electrons would be injected more easily and will move across the channel with less scattering, resulting in higher transconductance at relatively lower gate and drain biases in high-κ encapsulated TMD FETs.

To conclude, we have demonstrated that high-κ ATO films can be used as an n-type charge transfer dopant on ML $MoS_2$. The fact that ATO encapsulated ML $MoS_2$ devices exhibited comparable or better performance than previous doping and high-κ studies bears testimony to the superior doping and mobility enhancing capabilities of ATO thin films. Moreover, high-κ ATO can be deposited by a simple spin coating process which makes this doping approach attractive when compared to other time consuming doping techniques. Utilizing this technique on ML $MoS_2$, we demonstrated two-point field effect mobility as high as 83 $cm^2$/V-s at RT, four-point intrinsic mobility as high as 102 $cm^2$/V-s at RT and 501 $cm^2$/V-s at 77 K. ON-currents as high as 240 µA/µm for a 450nm channel length device, and a record low $R_C$ of 180 Ω·µm were demonstrated on ML MoS$_2$ after ATO encapsulation. In addition, we also shed light on the interfacial-oxygen-vacancy mediated doping of MoS$_2$ by high-κ dielectrics, in general, leading to improved screening of charged impurities, suppression of homopolar phonon scattering and reduction of the effective Schottky barriers at the contacts. Future work includes studying the stoichiometry and thickness scalability of ATO films and their effect on the performance and air stability of TMD-based devices.

*Conflict of Interest*: The authors declare no competing financial interest.

*Acknowledgment*: This work was supported by the NRI SWAN center, NSF NNIN, and Intel. The authors would like to thank the Army Research Office (ARO) for partial support of this work under STTR award number W911NF-14-P-0030.

*Supporting Information Available:* Description of materials and device fabrication methods, characterization tools and techniques, preparation of the ATO sol-gel precursor solution, XPS analysis of the as-deposited ATO films, performance degradation in ATO encapsulated devices and possible remedies, transfer characteristics temperature dependence and output characteristics at 77 K of a back-gated ML MoS$_2$ FET before/after ATO doping, deposition method of TiO$_2$ and its effect on ML MoS$_2$ FET performance, and transfer curves used for the mobility extraction of the two-point device in Figure 5(a). This material is available free of charge via the Internet at http://pubs.acs.org.

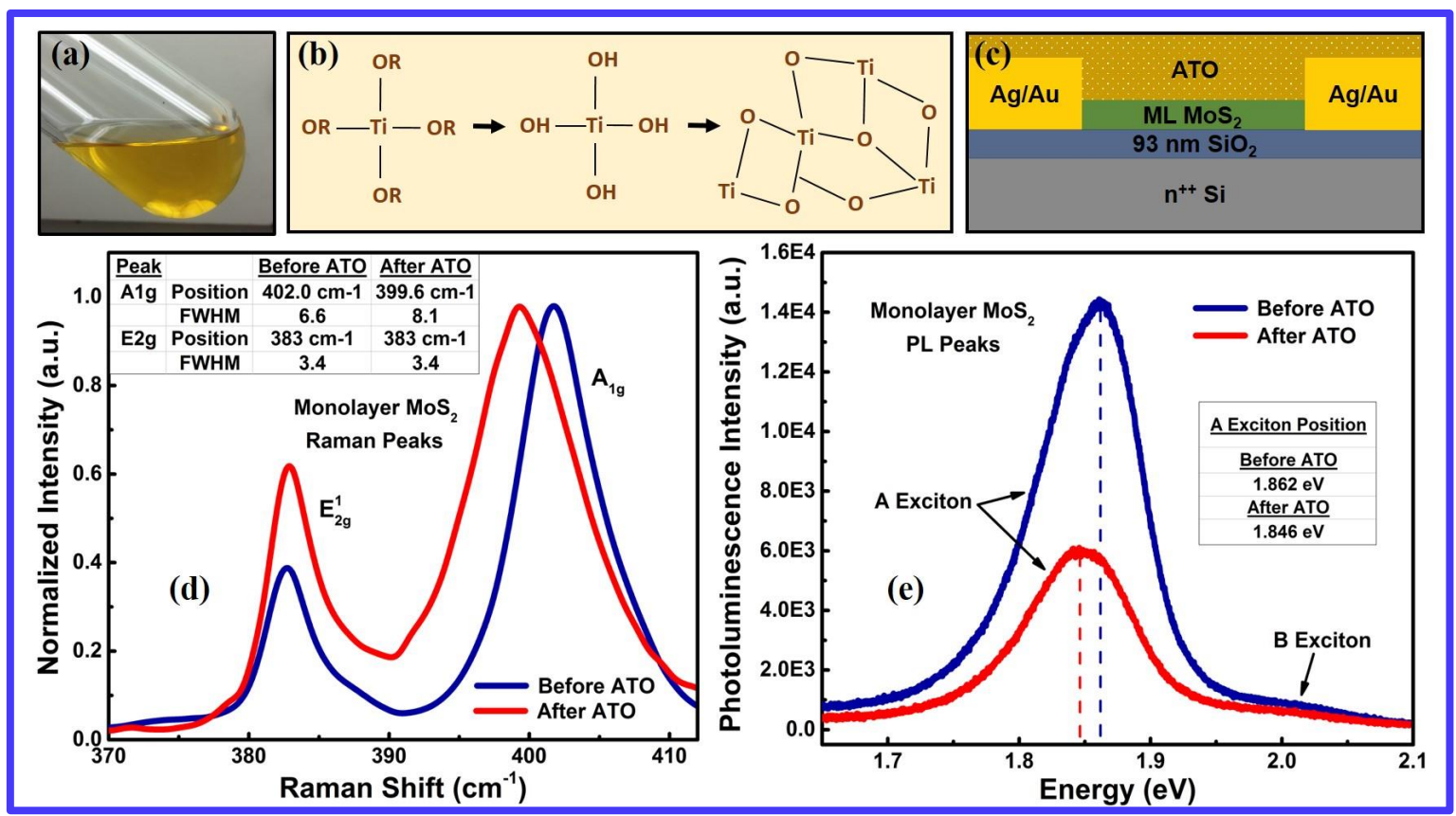

**Figure 1.** (a) Optical image of the as-prepared ATO precursor solution showing its characteristic yellowish orange color. (b) Schematic of chemical steps involved in the formation of ATO from its precursor molecules, namely titanium isopropoxide (R = CH (CH$_3$)$_2$). (c) Schematic of a representative back-gated ML MoS$_2$ FET encapsulated by ATO. (d) Raman spectra of ML MoS$_2$ showing its characteristic A$_{1g}$ and E$_{2g}^1$ peaks before (blue) and after (red) ATO encapsulation illustrating the electron doping-induced changes. (e) Photoluminescence spectra of ML MoS$_2$ before (blue) and after (red) ATO encapsulation showing a redshift in the peak position of the A exciton.

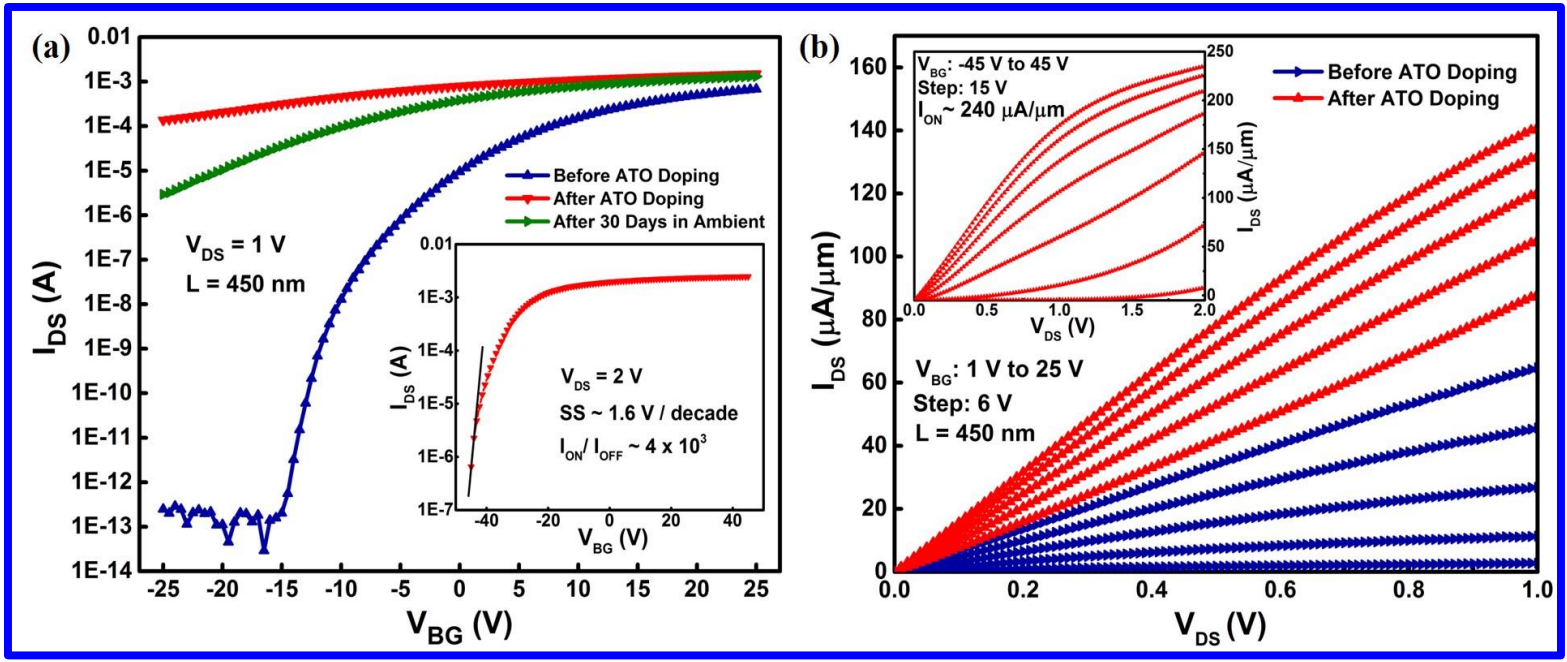

**Figure 2.** **(a)** Transfer characteristics, shown on a semi-log scale, of a representative ML MoS$_2$ FET at $V_{DS}$ = 1 V before (blue) and after (red) ATO doping, and after 30 days of ambient exposure (green). Inset shows the transfer characteristics of the doped FET measured under larger gate (-45 V to 45 V) and drain biasing (2 V) conditions. The channel length and width are 450 nm and 10.4 µm, respectively. **(b)** Output characteristics of the same FET before (blue) and after (red) ATO doping. Inset shows the output characteristics under larger biasing conditions with the ON-current reaching up to 240 µA/µm at a $V_{BG}$ of 45 V and $V_{DS}$ of 2 V.

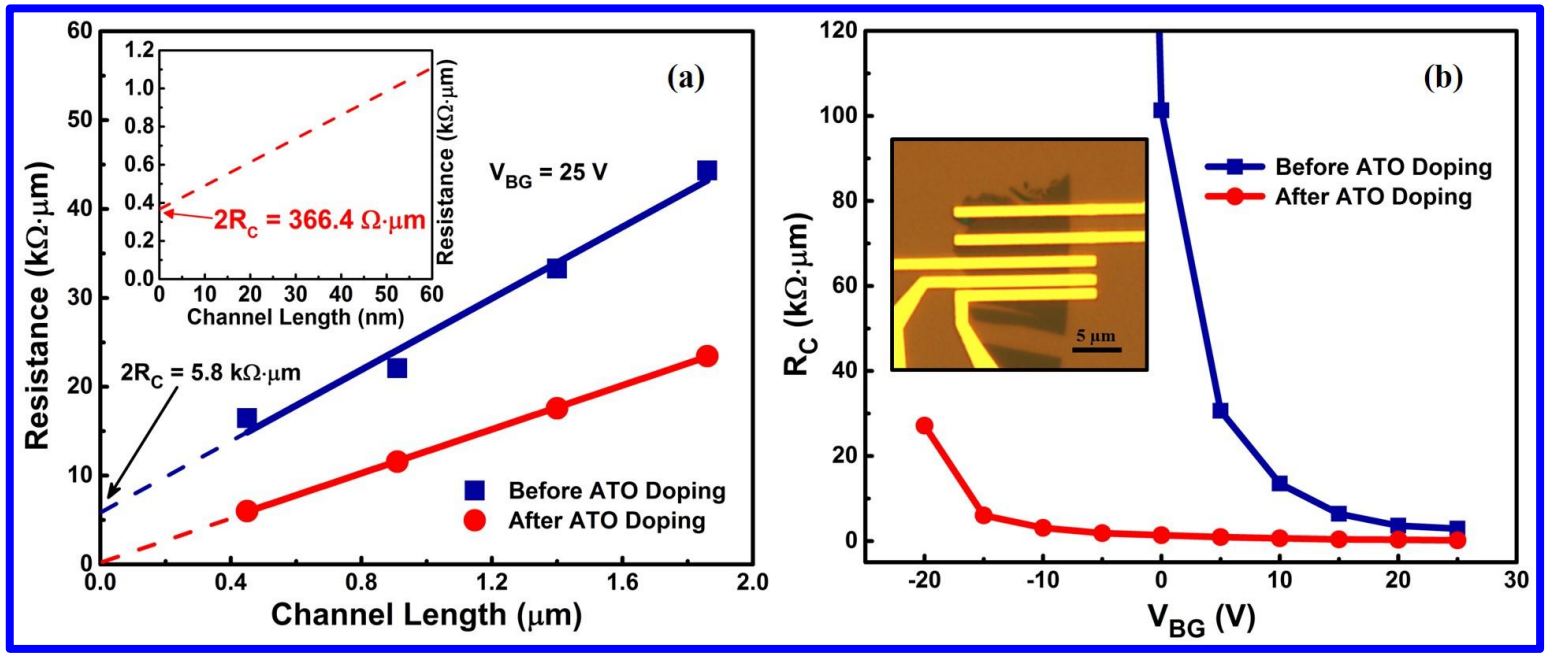

**Figure 3.** (a) Plot of total resistance as a function of channel length as determined from the TLM structure before (blue) and after (red) ATO doping at a $V_{BG}$ of 25 V. The solid blue and red lines are linear fits to the data. The $R_C$ and $L_T$ extracted before doping are 2.9 k$\Omega$·µm and 145 nm, respectively. After ATO doping, the extracted $R_C$ is ~ 180 $\Omega$·µm and $L_T$ is 15 nm. Inset: zoomed in view of the extrapolated dashed red line. (b) Extracted $R_C$ as a function of $V_{BG}$ before (blue) and after (red) ATO doping. The $R_C$ shows a strong gate dependence before doping and a weak gate dependence after doping. Inset: optical micrograph image of the as-fabricated TLM structure.

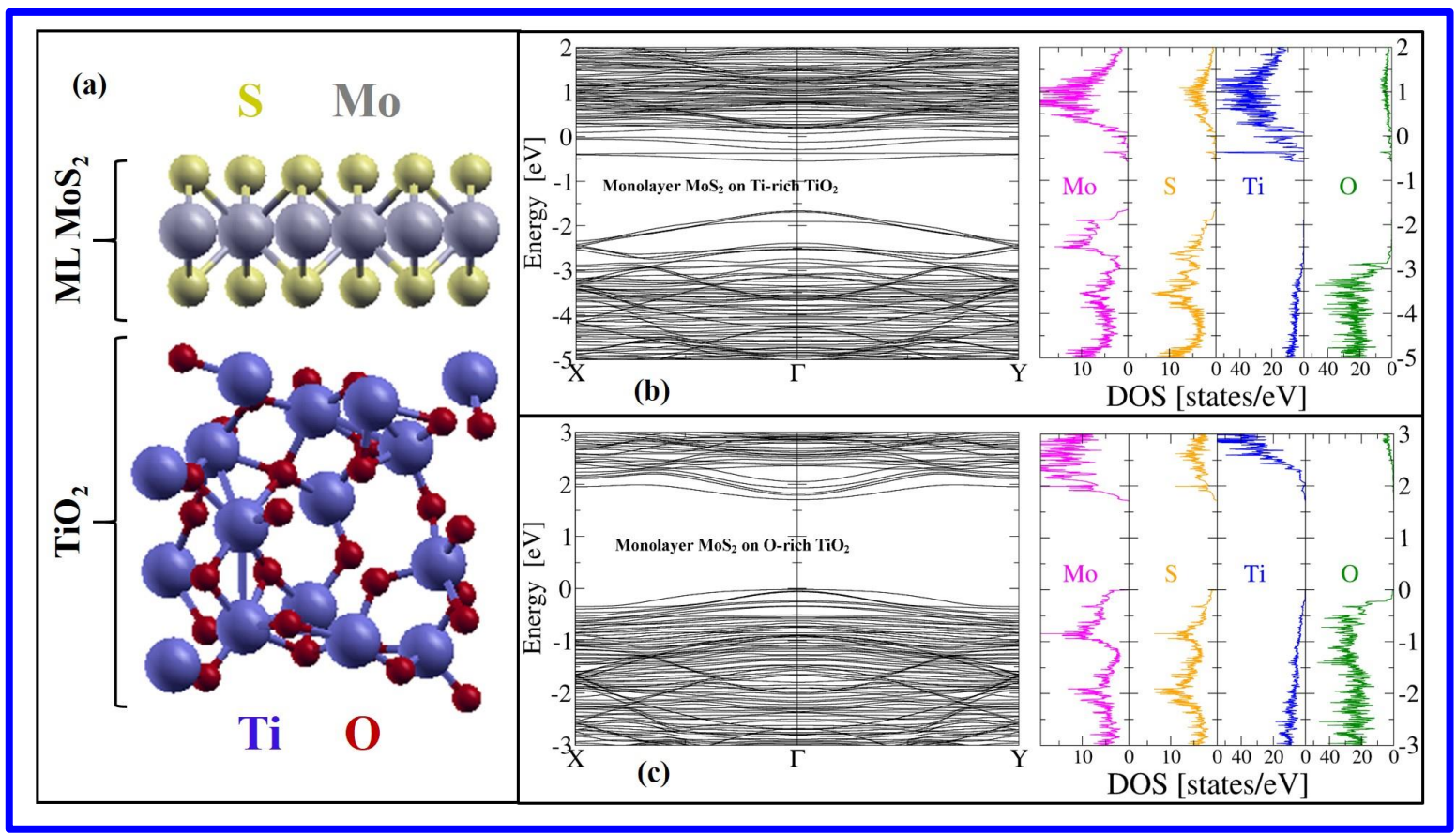

**Figure 4.** (a) Supercell showing the composite crystal structure consisting of ML MoS$_2$ on an underlying rutile-TiO$_2$ slab as simulated in VASP. For simulating the O-rich TiO$_2$ case, the TiO$_2$ slab was left unaltered. In contrast, for the Ti-rich TiO$_2$ case, a suitable number of O vacancies were created in TiO$_2$ at the ML MoS$_2$ – TiO$_2$ interface so as to mimic the MoS$_2$ – ATO scenario. (b) Band-structure and atom-projected-density-of-states (AP-DOS) plots for the ML MoS$_2$ – Ti-rich TiO$_2$ case. From the plots it can be deduced that in the presence of O vacancies, electronic states from Ti atoms are introduced near the conduction band edge of ML MoS$_2$ causing the Fermi level to get pinned above the conduction band indicating strong doping. (c) Band-structure and AP-DOS plots for the ML MoS$_2$ – O-rich TiO$_2$ case. No doping effect is seen in this case and the Fermi level remains pinned at the valence band edge. (Simulations were done assuming 0 K)

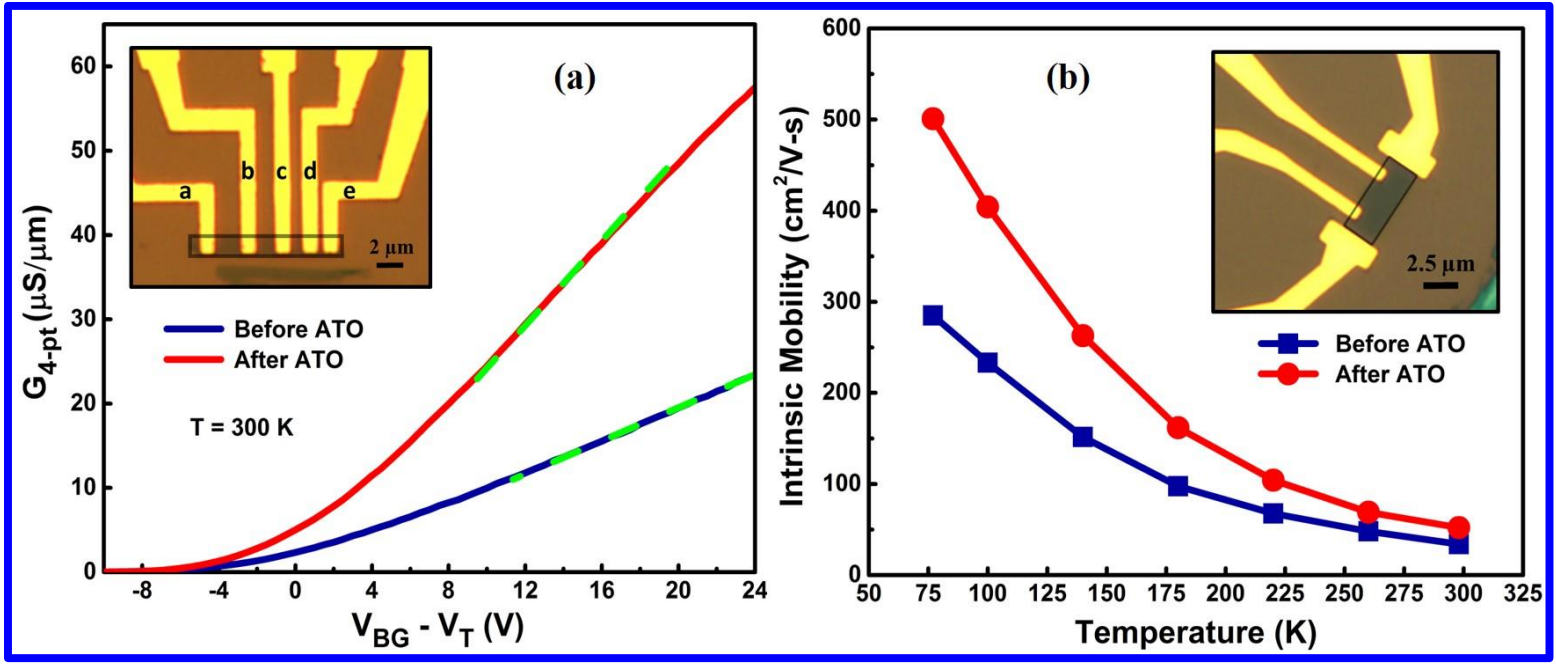

**Figure 5.** (a) Four-point conductance ($G_{4\text{-pt}}$) curves as a function of the gate overdrive ($V_{BG} - V_T$), measured between contacts a, b, c and d of the device shown in the inset, before (blue) and after (red) ATO encapsulation at RT. Dashed light green lines represent the regions from where the maximum slope was extracted for the calculation of intrinsic mobility of the ML MoS$_2$ flake before/after ATO encapsulation. The length and width of the active region are 1.45 µm and 1.42 µm, respectively. Contacts d and e (separated by 460 nm) were used to extract the two-point field effect mobility before/after ATO encapsulation. (b) Intrinsic mobility of ML MoS$_2$ as a function of temperature before (blue) and after (red) ATO encapsulation. Optical micrograph of the four-point device is shown in the inset. The length and width of the active region are 2.3 µm and 2.5 µm, respectively. The intrinsic mobilities are enhanced after ATO encapsulation reaching up to 501 cm$^2$/V-s at 77 K. (ML MoS$_2$ flakes are outlined at their edges)

# TOC Figure

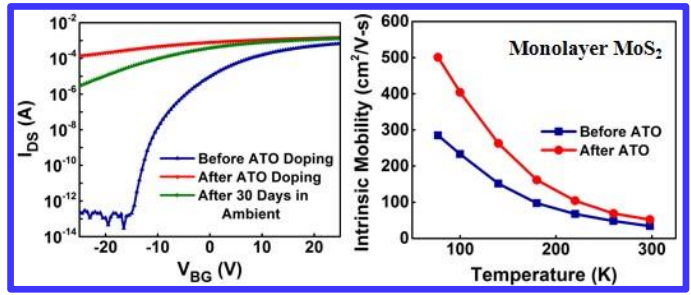

# Supporting Information

Air Stable Doping and Intrinsic Mobility Enhancement in Monolayer Molybdenum Disulfide by Amorphous Titanium Suboxide Encapsulation

Amritesh Rai[1,*], Amithraj Valsaraj[1], Hema C.P. Movva[1], Anupam Roy[1], Rudresh Ghosh[1], Sushant Sonde[1], Sangwoo Kang[1], Jiwon Chang[2], Tanuj Trivedi[1], Rik Dey[1], Samaresh Guchhait[1], Stefano Larentis[1], Leonard F. Register[1], Emanuel Tutuc[1] and Sanjay K. Banerjee[1]

[1] Microelectronics Research Center & Department of Electrical and Computer Engineering, The University of Texas at Austin, Austin, Texas 78758, USA

[2] SEMATECH, 257 Fuller Rd #2200, Albany, NY-12203, USA

[*] Correspondence e-mail: amritesh557@utexas.edu

# S1: Materials and Device Fabrication Methods

MoS$_2$ flakes were mechanically exfoliated, using the conventional 'scotch-tape' method, from a bulk MoS$_2$ crystal (SPI Supplies) onto degenerately doped ($\rho < 0.005$ $\Omega$-cm) n-type Si substrates covered with 93 nm thermally grown SiO$_2$. The oxide thickness was verified via ellipsometry measurements. Upon exfoliation, the samples were annealed in high vacuum (2 x 10$^{-6}$ Torr) at 350$^0$ C for 8 hours. This high vacuum annealing step helps minimize tape residues from the top surface of the flakes as well as trapped adsorbates, such as moisture, from in between the flake and the underlying SiO$_2$ substrate. A combination of optical contrast, atomic force microscope (AFM), Raman and photoluminescence (PL) measurements were used to identify atomically flat monolayer MoS$_2$ flakes of interest. Contacts on the flakes were patterned using standard electron beam lithography utilizing PMMA as the e-beam resist, followed by development in 1:3 MIBK:IPA to open up the pads. Electron beam evaporation (at a base pressure of 5 x 10$^{-6}$ Torr) and acetone lift-off steps were used to deposit a 20/30 nm stack of silver/gold (Ag/Au) which served as the contact electrodes. Ag was chosen as the contact metal due to the superior interface quality that it forms with MoS$_2$[1] besides having good adhesion with the SiO$_2$ substrate. No annealing was done after contact deposition. The contact width was fixed at 1µm. ATO thin films were deposited on top of the MoS$_2$ devices by spin-coating an ATO sol-gel precursor solution (85 mg/ml) followed by a short baking step, all of which was done in ambient conditions. The typical spin speed was 3000 rpm for a duration of 45 seconds, following which the samples were baked at 90$^0$ C on a hot plate for 15 minutes in order to dry the residual solvent and enable the conversion of the ATO precursor molecules into ATO through hydrolysis. The thickness of ATO films deposited in this manner was ~ 140 nm with an average surface roughness below 0.5 nm as determined from AFM. The $\kappa$ value of the ATO film was extracted to be ~ 10 from capacitance-voltage measurements.

# S2: Characterization Tools and Techniques

Optical investigation was done using an Olympus BX51M Microscope using their proprietary Stream Essentials analysis software. Ellipsometry measurements were taken using a JA Woollam M-2000 ellipsometer. Raman spectroscopy measurements were taken with a Renishaw inVia micro-Raman system with an excitation wavelength of 532 nm and a grating of 3000 l/mm. Photoluminescence measurements were taken with a Renishaw inVia micro-Raman system configured for photoluminescence with specialized optics at an excitation wavelength of 532 nm and a grating of 1200 lines/mm to obtain high energy peaks. Atomic force microscopy images were taken with a Veeco Nanoscope 5 in tapping mode. X-ray Photoelectron Spectroscopy was performed in a MULTIPROBE system from Omicron NanoTechnology GmbH utilizing a monochromatic Al-Kα source. Electrical characterization of the devices was done in dark using the Agilent 4156C and B1500A Semiconductor Parameter Analyzers. Ambient measurements were carried out in a Cascade Summit 11000 AP probe station. Low temperature and vacuum measurements ($< 5 \times 10^{-5}$ Torr) were carried out in a Lakeshore Cryotronics cryogenic probe station. All electrical measurements prior to ATO encapsulation were performed in vacuum in order to exclude the degrading effects of atmospheric adsorbates on the $MoS_2$ channel. All measurements post encapsulation were performed in ambient conditions except the low temperature measurements which were done in vacuum.

# S3: Preparation of the ATO Sol-Gel Precursor Solution

The ATO precursor solution was prepared utilizing the recipes as outlined in previous literature reports[2, 3]. The sol-gel preparation procedure is as follows: 25 ml of 2-methoxyethanol ($CH_3OCH_2CH_2OH$, Aldrich, 99.9+ %) and 2.5 ml of ethanolamine ($H_2NCH_2CH_2OH$, Aldrich, 99.0+ %) were first mixed in a cylindrical glass vessel equipped with a thermometer. The mixture was left for 10 minutes under magnetic stirring following which 5 ml of titanium (IV) isopropoxide (Ti $[OCH(CH_3)_2]_4$, Aldrich, 99.999 %) was added to the mixture. The cylindrical vessel containing the final mixture was then placed in a silicone-oil bath and was heated to 80 °C for a period of 2 hours under magnetic stirring. The temperature was then raised to 120 °C for 1 hour. This two-step heating cycle (80 °C – 2 hours + 120 °C – 1 hour) was then repeated a second time at the end of which the color of the solution turned yellowish orange, indicating the formation of the ATO precursor solution. The concentration of the as-prepared solution was determined to be ~ 85 mg/ml. This was done by completely evaporating the solvent from 5 mL of the as-prepared solution and measuring the weight of the residual crystallites. In the case of graphene, diluted solutions of the ATO precursor (10 mg/ml or 20 mg/ml) were used[3]. However, for doping the ML $MoS_2$, the as-prepared precursor solution with the high initial concentration was chosen because, unlike graphene, monolayer $MoS_2$ has a large band-gap and would need substantial doping to achieve high carrier densities unlike graphene.

# S4: X-Ray Photoelectron Spectroscopy (XPS) Analysis of ATO films

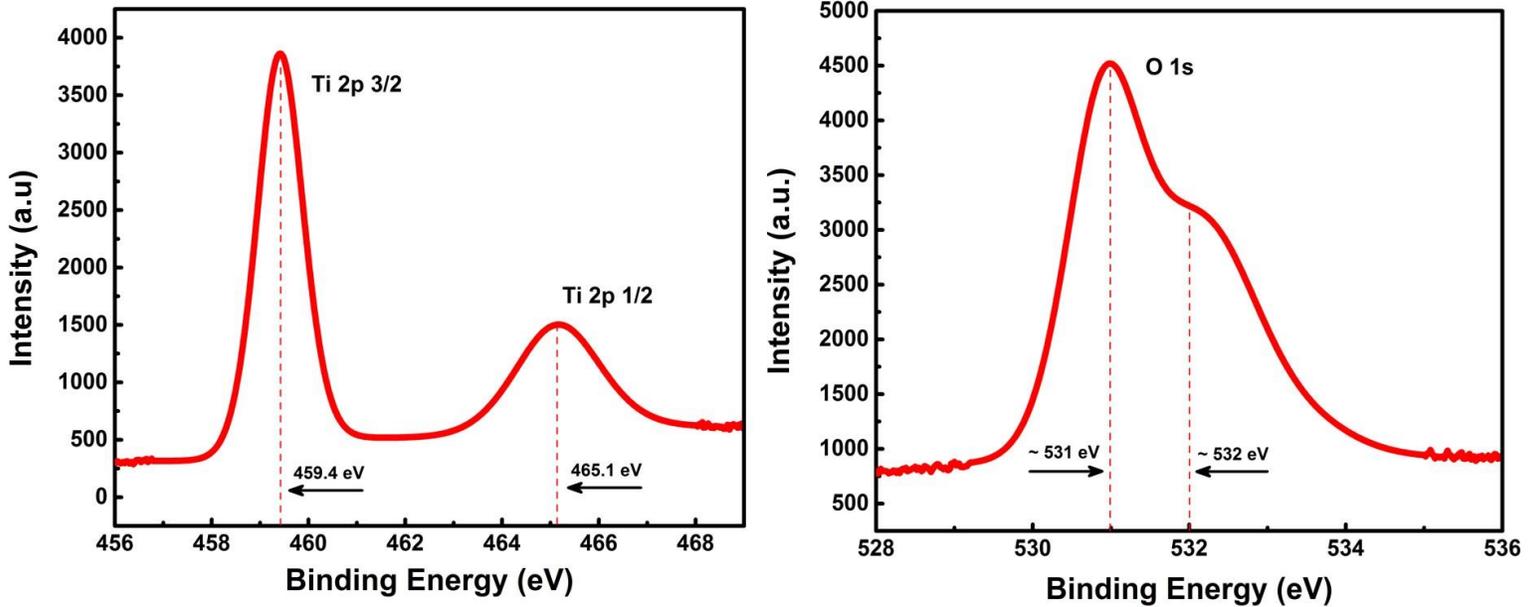

Figure S4: XPS spectra obtained from as-formed ATO films showing the Ti and O bonding states

Figure S4 above shows the XPS spectra as measured from the surface of our as-formed ATO films showing the corresponding binding energies of the Ti 2p 3/2, Ti 2p 1/2 and O 1s states. The elemental composition of our ATO film was determined by integrating the peak areas of the Ti 2p and O 1s spectra by properly fitting the components to each peak. It is to be noted that the O 1s peak shows the presence of two components. The peak at ~ 532 eV represents the un-bonded component of O probably resulting from OH species and, hence, it was not considered in our ratio determination. Only the shifted O 1s component at ~ 531 eV was considered as it represents bonding between the O and Ti atoms. We found the Ti:O ratio in our ATO films to be ~ 1:1.5.

# S5: Performance Degradation in ATO Encapsulated Devices

As discussed in the main manuscript, the doping effect observed in ATO encapsulated devices is absent when it is replaced by stoichiometric $TiO_2$. Therefore, it is reasonable to assume that the slight degradation observed in the performance of ATO-encapsulated $MoS_2$ devices over long term air exposure (30 days) could be due to the ATO becoming more O-rich at the ATO – $MoS_2$ interface owing to its interaction with the pre-adsorbed oxygen and water molecules on the strongly hydrophilic $SiO_2$ substrate. The resulting oxide or hydroxide formation can adversely impact the electron donating capability of Ti atoms to $MoS_2$ at the ATO – $MoS_2$ interface. Moreover, the pre-adsorbed oxygen or water molecules on the underlying $SiO_2$ substrate can react over time and degrade the quality of the ML $MoS_2$ itself. A possible way to eliminate any degradation effects could be to encapsulate the devices in ATO from both the top and bottom, especially since it has been shown that ATO is much more hydrophobic than $SiO_2$[2]. Other ways could be to use alternate substrates instead of $SiO_2$ such as h-BN. Moreover, further optimization of the preparation and deposition methods of the as-prepared ATO precursor solution is needed in order to minimize any impurities or trapped moisture in the overlaying dielectric.

# S6: Transfer Characteristics Temperature Dependence & Output Characteristics at 77 K of a Back-Gated ML MoS$_2$ FET Before/After ATO

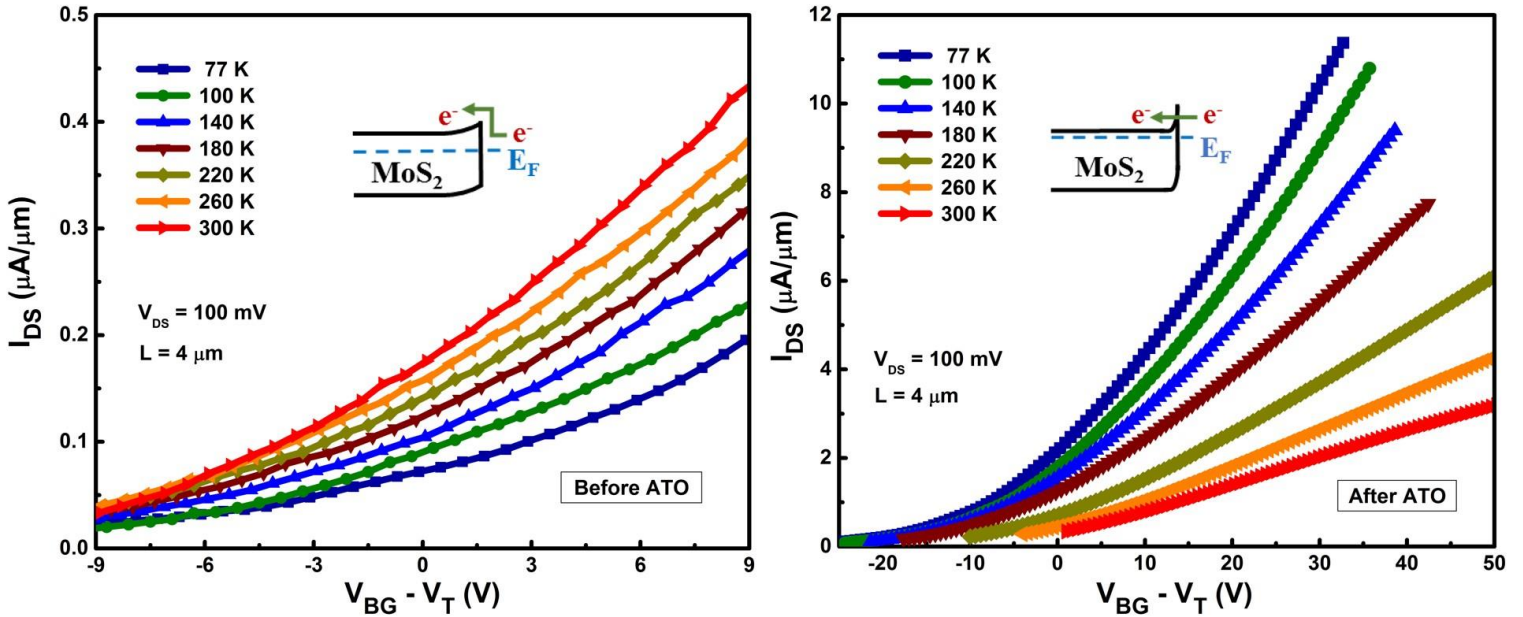

Figure S6.1: Temperature dependent transfer characteristics of a back-gated ML MoS$_2$ FET before/after ATO encapsulation

Figure S6.1 above shows the temperature dependent transfer characteristics of a back-gated ML MoS$_2$ FET (L = 4 µm, W = 2 µm) at a $V_{DS}$ of 100 mV. The x-axis is back-gate overdrive voltage ($V_{BG} - V_T$) and the $V_T$ of each individual curve was taken into account in generating the above plots. Before ATO doping (plot on the left), the current at a fixed gate overdrive voltage decreases as the temperature is lowered indicating Schottky-barrier limited transport which is dominated by thermionic emission over the barriers. After ATO doping (plot on the right), the trend reverses and the current at a fixed gate overdrive voltage increases as the temperature is lowered which is characteristic of phonon-limited transport. The dominant transport mechanism is no longer thermionic emission, but tunneling through the barriers as a consequence of doping-induced Schottky barrier width thinning.

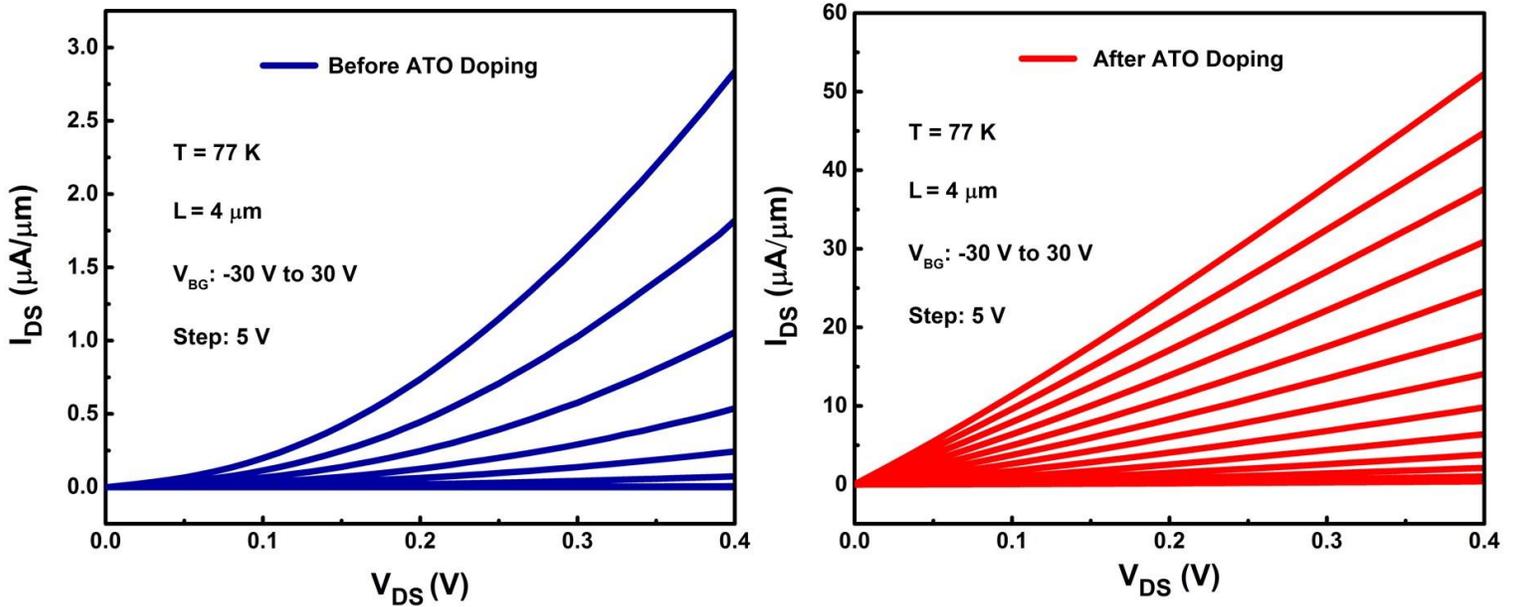

Figure S6.2: Output characteristics of the back-gated ML MoS$_2$ FET, presented above in Figure S6.1, at 77 K before/after ATO encapsulation

Figure S6.2 above shows the output characteristics of the back-gated ML MoS$_2$ FET, presented in Figure S6.1, measured at a temperature of 77 K before and after ATO doping. The effect of Schottky barriers on electron transport will be greater at 77 K owing to the reduced thermal energy of the carriers. The plot on the left (blue curves) depicts the bare MoS$_2$ FET clearly illustrating the exponential $I_{DS} - V_{DS}$ behavior indicative of substantial Schottky barriers between the MoS$_2$ and the Ag contact. In contrast, after the device is encapsulated in ATO, the output characteristics show a linear transport behavior indicating Ohmic contacts as depicted in the plot on the right (red curves). This linear behavior results due to the doping-induced thinning of the Schottky barrier width, thereby allowing the electrons to easily tunnel through.

## S7: Deposition Method of TiO$_2$ and its Effect on MoS$_2$ FET Performance

The method to deposit TiO$_2$ on MoS$_2$ was adopted from a recent report of forming TiO$_2$ dielectrics on graphene as demonstrated by Corbet *et al.*[4] Using an SEC-600 e-beam evaporator from CHA Industries, high purity titanium pellets were evaporated from a titanium carbide crucible at a base pressure of 5 x 10$^{-6}$ Torr which further reduced to 1 x 10$^{-6}$ Torr during Ti deposition. Ti films with a thickness of 1 nm were evaporated at a rate of < 0.1 Å/s with the chamber being vented to atmosphere after each 1 nm of deposition in order to oxidize the Ti film to TiO$_2$. A 5 cycle deposition was performed which resulted in a TiO$_2$ film about 6 nm thick as has been demonstrated using ellipsometry and TEM analysis[4]. Furthermore, XPS measurements reported on titanium oxide deposited in this manner revealed a pure TiO$_2$ film[4].

Unlike the ATO films, the stoichiometric TiO$_2$ film did not show any doping effect when deposited on back-gated ML MoS$_2$ FETs. Instead, as illustrated in Figure S6 below, the device showed a performance degradation. This can be explained by the poor interface quality that probably forms between MoS$_2$ and PVD deposited TiO$_2$. In fact, mobility degradation was also reported in graphene FETs with TiO$_2$ gate dielectrics[4]. Furthermore, the degradation could also be due to short range scattering by TiO$_2$ impurities as has been observed in the case of graphene[5].

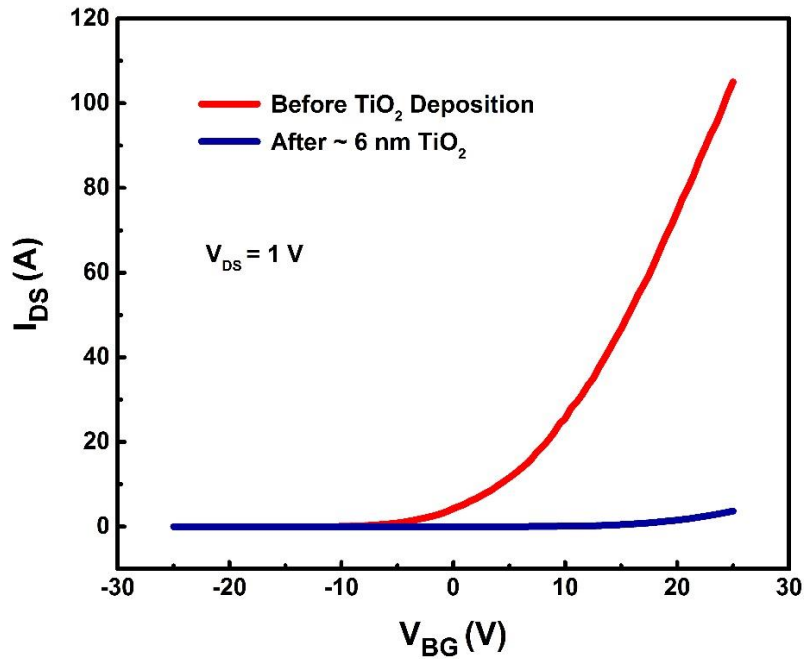

Figure S7: Transfer characteristics of a back-gated ML MoS$_2$ FET before (red) and after (blue) ~ 6 nm TiO$_2$ deposition at a V$_{DS}$ of 1 V

# S8: Transfer Characteristics of the FET presented in Figure 5(a)

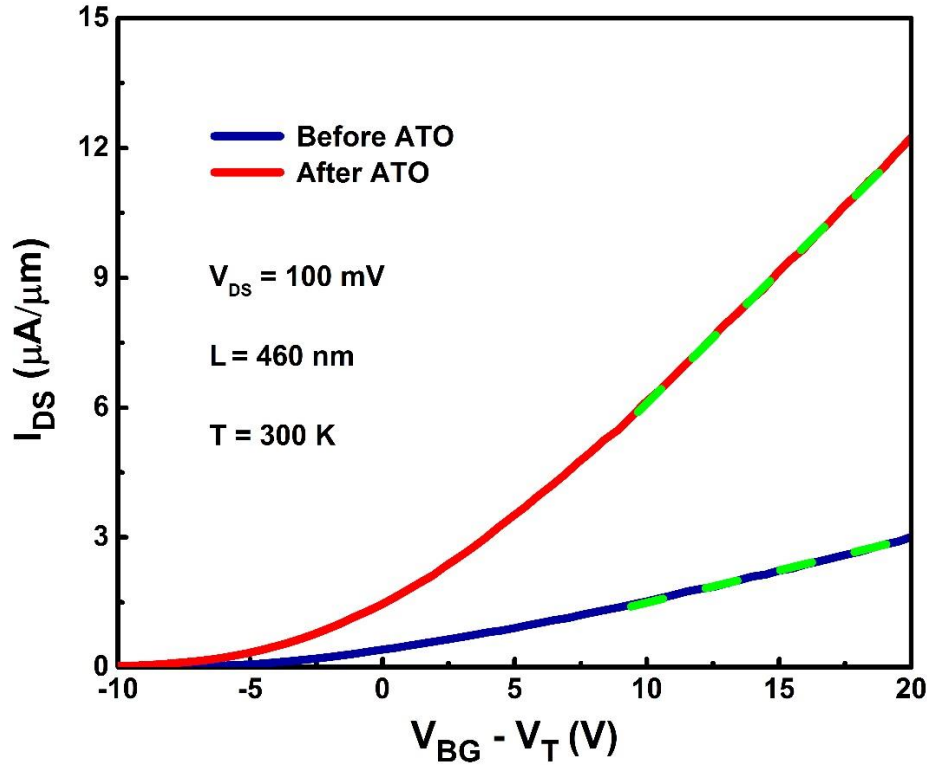

Figure S8: Transfer characteristics of the FET between contacts 'd' & 'e' of the multi-contact device shown in Figure 5(a) of the manuscript

Figure S8 above shows the transfer curves of the two-point FET between contacts 'd' & 'e' of Figure 5(a) at a $V_{DS}$ of 100 mV measured at RT before (blue) and after (red) ATO encapsulation. From the regions of maximum slope as marked in the figure by dashed light-green lines, the peak $g_m$ was extracted both before and after ATO doping. The $\mu_{FE}$ calculated before doping was ~ 24 cm$^2$/V-s which increased to ~ 83 cm$^2$/V-s after doping showing a > 3X improvement in the field effect mobility. This two-point device with a channel length of 460 nm showed the best mobility enhancement upon ATO encapsulation among all other two-point devices with a similar channel length.